\documentclass[useAMS,usenatbib]{mn2e}

\usepackage{graphicx}
\usepackage{amssymb}
\newcommand{\apj}{ApJ}

\newcommand{\aap}{A\&A}

\newcommand{\aj}{AJ}
\newcommand{\mnras}{MNRAS}

\newcommand{\pasp}{PASP}
\newcommand{\apjs}{ApJS}
\newcommand{\MC}{\multicolumn}
\newcommand{\kms}{km\,s$^{-1}$}
\newcommand{\Te}{T$_{\rm e}$}
\newcommand{\sunn}{$_{\odot}$}
\DeclareRobustCommand{\ion}[2]{%
\relax\ifmmode
\ifx\testbx\f
{\mathrm{#1\,\textsc{#2}}}\else
{\mathrm{#1\,\mathsc{#2}}}\fi
\else\textup{#1\,{\mdseries\textsc{#2}}}%
\fi}

\title[A massive Z=Z$_{\odot}$/36 star in DDO~68]{Discovery of a massive
variable star with Z=Z$_{\odot}$/36 in the galaxy DDO~68}
\author[S. A. Pustilnik, A. L. Tepliakova, A. Y. Kniazev, A. N. Burenkov]
{S. A. Pustilnik,$^1$\thanks{sap@sao.ru (SAP), arina@sao.ru (ALT), aknizev@saao.ac.za (AYK), ban@sao.ru (ANB)}
A. L. Tepliakova,$^1$\footnotemark[1]
A. Y. Kniazev,$^{2,1}$\footnotemark[1]
A. N. Burenkov$^1$\footnotemark[1]\\
$^1$ Special Astrophysical Observatory of RAS, Nizhnij Arkhyz, Karachai-Circassia 369167, Russia\\
$^2$ South African Astronomical Observatory, Observatory Road, Cape Town, South Africa}

\begin{document}

\label{firstpage}

\date{Accepted 2008 May 1. Received 2008 April 26}

\pagerange{\pageref{firstpage}--\pageref{lastpage}} \pubyear{2008}

\maketitle

\begin{abstract}

The Local Volume dwarf galaxy DDO~68, from the spectroscopy of its two
brightest HII regions (Knots 1 and 2) was designated as the second most
metal-poor star-forming galaxy [12+$\log$(O/H)=7.14].
In the repeated spectral observations in 2008 January with the 6-m telescope
(BTA) of the HII region Knot 3 [having 12+$\log$(O/H)=7.10$\pm$0.06],
we find a strong evidence of a transient event related to a massive star
evolution. From the follow-up observation with the higher spectral resolution
in 2008 February, we confirm this phenomenon, and give parameters of its
emission-line spectrum comprising of Balmer HI and \ion{He}{i} lines.
The luminosities of the strongest transient
lines (H$\alpha$, H$\beta$) are of a few 10$^{36}$~erg~s$^{-1}$.
We also detected an additional continuum component in the new spectrum of
Knot~3, which displays the spectral energy distribution raising to
ultraviolet. The estimate of the flux of this continuum leads us to its
absolute $V$-band magnitude of $\sim$--7.1.
Based on the spectral properties of this transient component, we suggest that
it is related to an evolved massive star of luminous blue variable type
with Z=Z$_{\odot}$/36.
We briefly discuss observational constraints on parameters of this unique
(in the aspect of
the record low metallicity of the progenitor massive star) event and propose
several lines of its study.

\end{abstract}


\section[]{INTRODUCTION}
\label{sec:intro}

The most metal-poor galaxies [or extremely metal-deficient, XMD,
with 12+$\log$(O/H)$<$7.65, e.g., review by \citet{KO}]
are considered as the best laboratories to study in great detail
the processes related to galaxy evolution and star formation (SF)
in young galaxies at high redshifts. Among several XMD galaxies with
O/H near the bottom of dwarf galaxy distribution, namely those with
parameter 12+$\log$(O/H)$<$7.3, DDO~68 (UGC~5340) with its value of 7.14
is the nearest one \citep[][ hereafter PKP and IT07]{DDO68,IT07}.
Its distance was
accepted as 6.5 Mpc in PKP. However, for its V$_{\rm helio}$=502~\kms, due
to the large negative peculiar velocity correction
in this space region \citep{Tully08}, this can be as large
as $\sim$10 Mpc. Nevertheless, this galaxy is one of the most perspective
to study individual massive stars with the future
Extremely Large Telescopes (ELTs) \citep[e.g.,][]{IAUS232}. In the recent
papers DDO~68
was shown to be a likely result of a merger of two gas-rich dwarfs
\citep*{Ekta08} and to show up colours on the SDSS images,
which for the standard Salpeter IMF are consistent with no tracers of stars
older than $\sim$1~Gyr \citep*{DDO68_sdss}.
The current O/H value of DDO~68 is based on the direct measurements by
the classic \Te-method of two brightest HII regions in the Northern ring,
called Knots 1 and 2 (PKP, IT07). In this Letter we present new BTA
spectra of Knots 1 and 3. The latter displays transient emission lines
and blue continuum, which we attribute to a massive variable star of
luminous blue variable (LBV) type.

\section[]{OBSERVATIONS AND DATA REDUCTION}
\label{sec:obs}

The long-slit spectral observations of Knots 1 and 3 (15\arcsec\ in between)
were conducted with the multimode instrument SCORPIO \citep{SCORPIO}
installed in the prime focus of the SAO 6\,m telescope (BTA) on the
nights of 2008 January 11 and February 4.
The grism VPHG550G was used with the 2K$\times$2K CCD detector EEV~42-40
on 2008 January 11 and the grism VPHG1200G on 2008 February 4. These set-ups
gave the range 3500--7500~\AA\ with $\sim$2.1~\AA~pixel$^{-1}$ and the
full width at half-maximum (FWHM) of $\sim$12~\AA\ and the range
3900--5700~\AA\ with $\sim$0.9~\AA~pixel$^{-1}$ and FWHM of $\sim$6~\AA\
along the dispersion, respectively.
The scale along the slit (after binning) was 0\farcs36 pixel$^{-1}$.
Six and five 15-min. exposures were obtained on nights 2008 January 11 and
2008 February~4, respectively, under the  seeing of 1.2 and
1.3 arcsec.
The objects spectra were complemented before or after by the reference
spectra of He--Ne--Ar lamp for the wavelength calibration.
Bias and flat-field images were also acquired to perform the standard
reduction of 2D spectra. Spectral standard star Feige~34
\citep{Bohlin96} was observed during the night for the flux
calibration.

All data reduction for Knots 1 and 3 and emission line measurements for
Knot 1 were performed similar to that described in PKP. Namely
the standard pipeline with the use of IRAF\footnote{IRAF: the Image Reduction
and Analysis Facility is distributed by the National Optical Astronomy
Observatory, which is
operated by the Association of Universities for Research in Astronomy,
In. (AURA) under cooperative agreement with the National Science
Foundation (NSF).}
and MIDAS\footnote{MIDAS is an acronym for the European Southern
Observatory package -- Munich Image Data Analysis System. }
was applied for the reduction of long-slit spectra, which included the next
steps:
cosmic ray hits removal, bias subtraction,  flat-field correction, wavelength
calibration, night-sky background subtraction. Then, using the data on
spectrophotometry standard stars, all spectra were transformed to absolute
fluxes.
1D spectra were extracted by summing up, without weighting,
of 12 and 6 rows along the slit, in Knots 1 and 3, respectively.
The emission lines with their errors were measured in the way described
in details in \citet{SHOC}.

\section[]{RESULTS}
\label{sec:results}

\begin{figure}
  \centering
 \includegraphics[angle=-90,width=7.5cm, clip=]{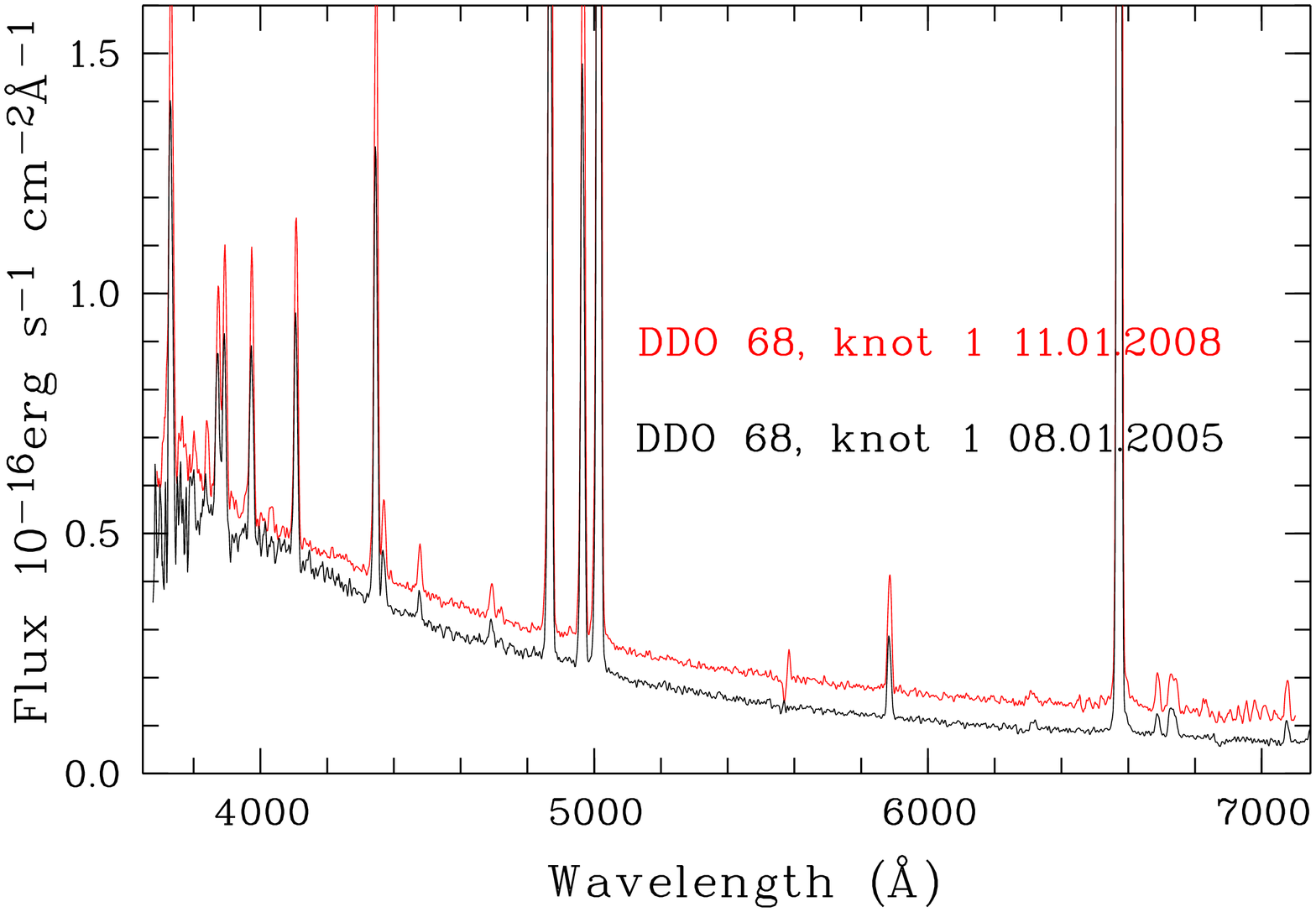}
 \includegraphics[angle=-90,width=7.5cm, clip=]{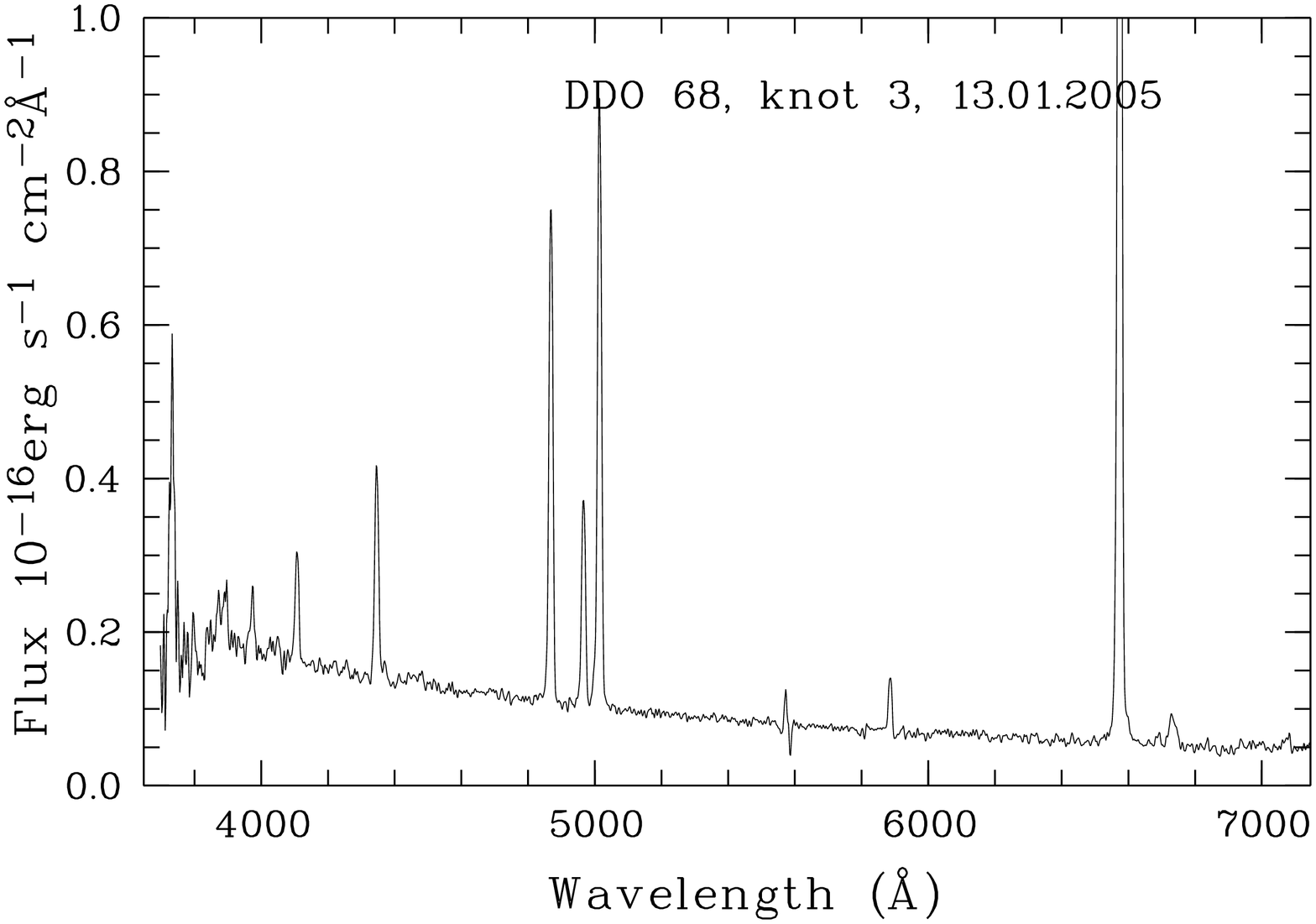}
 \includegraphics[angle=-90,width=7.5cm, clip=]{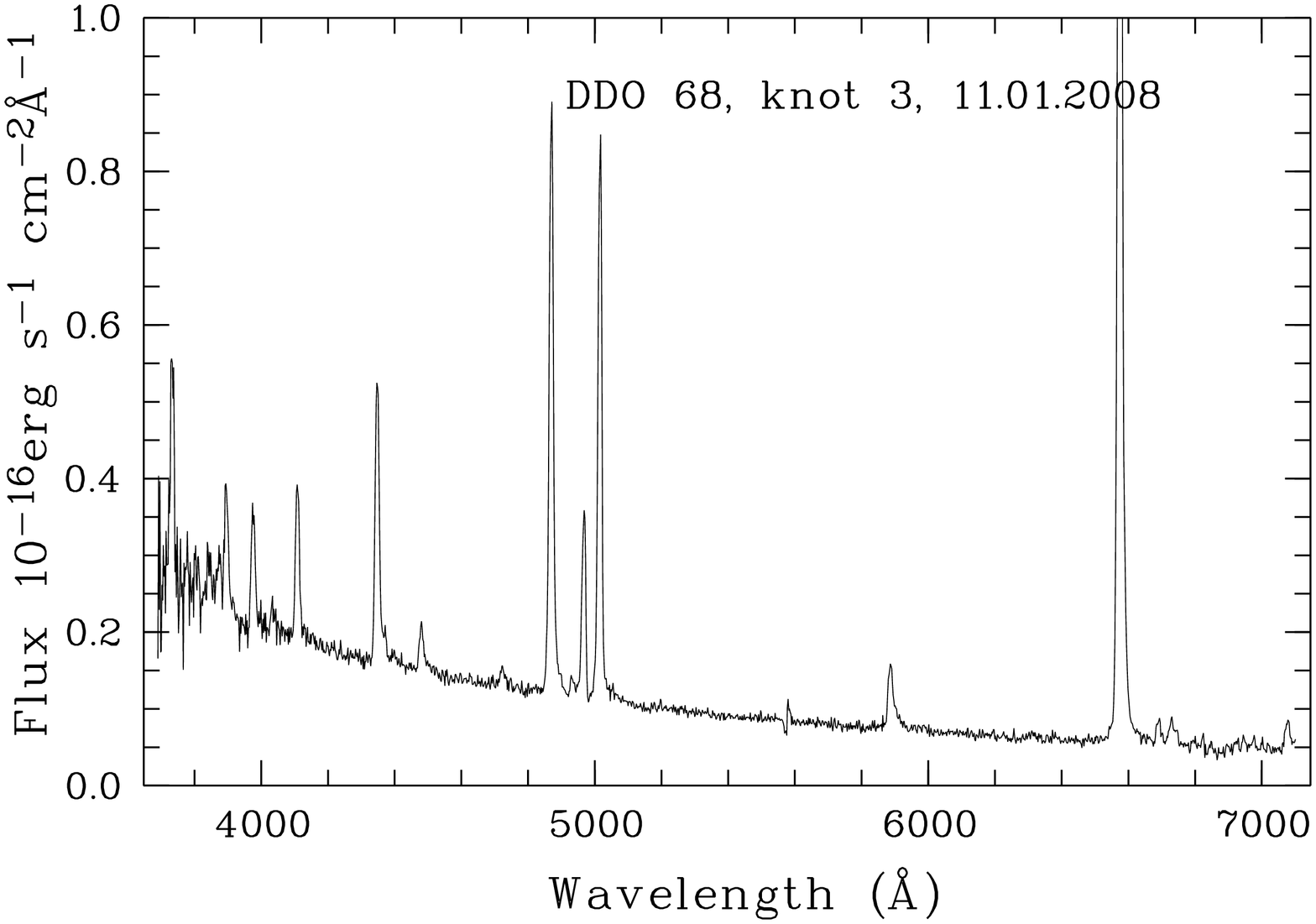}
  \caption{
{\bf Top panel}: The spectra of DDO~68 Knot 1 on 2005 January 8 and  2008
January 11, superimposed, with the former one shifted down in flux by 0.05
 unit. The spectra look very similar, including both lines and continuum.
{\bf Middle panel}: Spectrum of Knot 3 in DDO~68 on 2005 January 13.
{\bf Bottom panel}: Spectrum of Knot 3 in DDO~68 on 2008 January 11.
 Pay attention on the significant enhancement in the latter spectrum
 of \ion{He}{i} lines ($\lambda$4471, $\lambda$4713, $\lambda$5876,
  $\lambda$6678 and $\lambda$7065), the appearance of asymmetric broad
  wings in the
 strongest \ion{He}{i} lines, and the change of relative intensities
 of [\ion{O}{iii}]$\lambda$4959,5007 and Balmer lines.
}
	\label{fig:knot3}
 \end{figure}

In Fig.~\ref{fig:knot3} the relevant new and old spectra, obtained with
the same set-up, are shown.
The  bottom panel displays the spectrum of Knot 3 obtained on 2008 January
11 with
FWHM$\sim$12~\AA, while the middle panel shows the similar spectrum of Knot 3
obtained on 2005 January 13. Significant changes in relative line intensities
have occurred during the past $\sim$3~years. In the top panel, we show two
spectra of Knot 1, obtained on 2008 January 11  (upper) and 2005 January 8.
The latter
spectrum is shifted down by 0.05 flux unit to allow better visual comparison
of both spectra. They have very similar parameters of both relative line
intensities and underlying continuum and show no evidences for systematic
difference.

In Table~\ref{t:Knot1_new}, we present the line intensities of all relevant
emission lines measured in the spectrum of Knot 1 of 2008 January 11,
normalized
by the intensity of H$\beta$ and corrected for
the foreground extinction C(H$\beta$) and equivalent widths of underlying
Balmer absorption lines (EW(abs)),  given in the bottom part of this table.

\begin{table}
\centering{
\caption{Line intensities and derived parameters of Knot 1}
\label{t:Knot1_new}
\begin{tabular}{lcc} \hline  \hline
\rule{0pt}{10pt}
& \MC{2}{c}{PA=--26\degr,  $\theta$=1\farcs2}   \\ \cline{2-3}
\rule{0pt}{10pt}
$\lambda_{0}$(\AA) Ion & $F$($\lambda$)/$F$(H$\beta$) & $I$($\lambda$)/$I$(H$\beta$) \\ \hline

3727\ [O\ {\sc ii}]\            & 0.360$\pm$0.013 & 0.355$\pm$0.013    \\
3869\ [Ne\ {\sc iii}]\          & 0.152$\pm$0.006 & 0.150$\pm$0.006     \\
3889\ He\ {\sc i}\ +\ H8\       & 0.181$\pm$0.006 & 0.212$\pm$0.008     \\
3967\ [Ne\ {\sc iii}]\ +\ H7\   & 0.170$\pm$0.006 & 0.200$\pm$0.008     \\
4026\ He\ {\sc i}\              & 0.012$\pm$0.002 & 0.012$\pm$0.002    \\
4101\ H$\delta$\                & 0.226$\pm$0.007 & 0.251$\pm$0.008    \\
4340\ H$\gamma$\                & 0.466$\pm$0.013 & 0.482$\pm$0.014    \\
4363\ [O\ {\sc iii}]\           & 0.057$\pm$0.006 & 0.056$\pm$0.006     \\
4471\ He\ {\sc i}\              & 0.030$\pm$0.002 & 0.030$\pm$0.002   \\
4686\ He\ {\sc ii}\             & 0.022$\pm$0.002 & 0.022$\pm$0.002    \\
4713\ [Ar\ {\sc iv]}\ +\ He\ {\sc i}\ & 0.005$\pm$0.002 & 0.005$\pm$0.002   \\
4861\ H$\beta$\                 & 1.000$\pm$0.003 & 1.000$\pm$0.003    \\
4922\ He\ {\sc i}\              & 0.005$\pm$0.002 & 0.005$\pm$0.002    \\
4959\ [O\ {\sc iii}]\           & 0.610$\pm$0.020 & 0.601$\pm$0.020    \\
5007\ [O\ {\sc iii}]\           & 1.872$\pm$0.052 & 1.844$\pm$0.052    \\
5876\ He\ {\sc i}\              & 0.074$\pm$0.003 & 0.073$\pm$0.003    \\
6300\ [O\ {\sc i}]\             & 0.008$\pm$0.002 & 0.008$\pm$0.002    \\
6312\ [S\ {\sc iii}]\           & 0.006$\pm$0.002 & 0.006$\pm$0.002    \\
%6548\ [N\ {\sc ii}]\            & ...             & ...                \\
6563\ H$\alpha$\                & 2.708$\pm$0.070 & 2.672$\pm$0.077    \\
%6584\ [N\ {\sc ii}]\            & ...             & ...                \\
6678\ He\ {\sc i}\              & 0.026$\pm$0.002 & 0.026$\pm$0.002    \\
6717\ [S\ {\sc ii}]\            & 0.029$\pm$0.003 & 0.028$\pm$0.003    \\
6731\ [S\ {\sc ii}]\            & 0.025$\pm$0.003 & 0.025$\pm$0.003    \\
7065\ He\ {\sc i}\              & 0.019$\pm$0.002 & 0.018$\pm$0.002    \\
%7136\ [Ar\ {\sc iii}]\          & 0.026$\pm$0.004 & 0.023$\pm$0.004     \\
%
\hline
Parameter                 & \MC {2}{c}{Value}           \\ \hline
C(H$\beta$)\ dex          & \MC {2}{c}{0.00$\pm$0.03}   \\
EW(abs)\ \AA\             & \MC {2}{c}{2.50$\pm$0.14}   \\
$F$(H$\beta$)$^a$\        & \MC {2}{c}{39.1$\pm$0.7}     \\
EW(H$\beta$)\ \AA\        & \MC {2}{c}{ 162$\pm$3.9}     \\
Rad. vel.\ \kms\          & \MC {2}{c}{ 476$\pm$33}      \\
& \\
$T_{\rm e}$(\ion{O}{iii})(10$^{3}$~K)\     & \MC {2}{c}{18.81$\pm$1.12}      \\
$T_{\rm e}$(\ion{O}{ii})(10$^{3}$~K)\      & \MC {2}{c}{15.31$\pm$0.86}      \\
$T_{\rm e}$(SIII)(10$^{3}$~K)\             & \MC {2}{c}{17.32$\pm$0.93}      \\
$N_{\rm e}$(SII)(cm$^{-3}$)\               & \MC {2}{c}{378$\pm$294}~       \\
& \\
O$^{+}$/H$^{+}$($\times$10$^{-5}$)\        & \MC {2}{c}{0.301$\pm$0.044}     \\
O$^{++}$/H$^{+}$($\times$10$^{-5}$)\       & \MC {2}{c}{1.192$\pm$0.167}     \\
O$^{+++}$/H$^{+}$($\times$10$^{-5}$)\      & \MC {2}{c}{0.039$\pm$0.006}     \\
O/H($\times$10$^{-5}$)\                    & \MC {2}{c}{1.532$\pm$0.173}     \\
12+log(O/H)\                               & \MC {2}{c}{~7.19$\pm$0.05}~     \\
&   \\
Ne$^{++}$/H$^{+}$($\times$10$^{-5}$)\      & \MC {2}{c}{0.201$\pm$0.027}~~  \\
ICF(Ne)\                                   & \MC {2}{c}{1.285}              \\
log(Ne/O)\                                 & \MC {2}{c}{--0.77$\pm$0.08}~~  \\
&   \\
S$^{+}$/H$^{+}$($\times$10$^{-7}$)\        & \MC {2}{c}{0.527$\pm$0.058}~~  \\
S$^{++}$/H$^{+}$($\times$10$^{-7}$)\       & \MC {2}{c}{3.132$\pm$0.742}~~   \\
ICF(S)\                                    & \MC {2}{c}{1.601}             \\
log(S/O)\                                  & \MC {2}{c}{--1.56$\pm$0.13}~~  \\
\hline   \hline
\MC{3}{l}{$^a$ in units of 10$^{-16}$ ergs\ s$^{-1}$cm$^{-2}$.}\\
\end{tabular}
}
\end{table}

We also present in this table the electron temperatures for
zones of emission of [\ion{O}{iii}], [\ion{O}{ii}] and [\ion{S}{iii}],
N$_{\rm e}$, and ionic abundances of oxygen, neon and sulphur (with
respective ICFs - ionization correction factors), along with the total
abundances derived for the above measured line intensities with the classic
\Te\ method as described in \citet{SHOC,Sextans}. The latter are consistent
with the results from both PKP and IT07.
The relative line intensities of Knot~1, measured on the spectrum 2008
February 4
(not shown here due to the lack of space) with a higher resolution but with
a more narrow range,
are consistent with those from Table~\ref{t:Knot1_new}. However,
they can not be used for the independent derivation of O/H since this
spectrum does not include the line [O\ {\sc ii}]$\lambda$3727.

One of the aims of new observations of Knot~3 was an attempt to improve
its element abundances determination. However, due to appearance in the
spectrum of 20008 January 11 of additional `broad' components and the difficulty
due to low spectral resolution (FWHM$\sim$12~\AA) to disentangle
nebular emission lines from variable `broad' components, we should postpone
this task till getting the spectrum with a higher resolution. The same
relates to the spectrum of 2008 February 4, with
FWHM$\sim$6~\AA. This appears also insufficient to get reliable line
intensities of nebular emission lines.
Before to return to analysis of lines of the variable component,
we briefly describe one more estimate of O/H for Knot~3, derived
from its old spectrum, which will be used in Discussion.

From the line intensities of Knot~3 in table~3 and fig.~2 of PKP (also
shown in the middle panel of Fig.~\ref{fig:knot3}), with the use of direct
T$_{\rm e}$ method for the observation with the better S-to-N ratio in
I([\ion{O}{iii}$\lambda$4363]) (2005 January 13), the derived value
of 12+$\log$(O/H), assumed to be related to the metallicity of the variable
star, discussed below, was 7.11$\pm$0.11.
This motivates an attempt to improve the accuracy of O/H in Knot~3.

We apply below the so-called `semi-empirical' method, suggested by IT07.
It is intended for abundance determination in the case when the principal
faint line [\ion{O}{iii}]$\lambda$4363 is too noisy or undetected. This uses
the rather tight correlation between T(\ion{O}{iii}) and the total intensity
of
the strong oxygen lines relative to I(H$\beta$) (Stasi\'nska \& Izotov 2003).
After T(\ion{O}{iii}) is estimated from the total intensity of
[\ion{O}{ii}]$\lambda$3727, [\ion{O}{iii}]$\lambda\lambda$4959,5007 lines
via the
relation presented in IT07, all other calculations are made as in the classic
\Te\ method, using measured line intensities. The robustness of the
semi-empirical method in the very low-metallicity regime was tested in
IT07 and by us, using all available data on the DDO~68 HII regions. These O/H
estimates relative to those obtained by the direct \Te\ method, appeared very
stable and showed a scatter consistent with the errors of the direct method
determination and those for the semi-empirical method itself. Applying this
method to the same data on line intensities, and ignoring the faint line
[\ion{O}{iii}]$\lambda$4363, we obtain for Knot~3 the value of
12+$\log$(O/H)=7.10$\pm$0.06, which is nicely consistent with the value
derived from the direct \Te\ method.

We therefore accept for further discussion this value of O/H, which
corresponds to 1/36 of (O/H)\sunn\ \citep{solar04}.
As the nearest of young star-forming regions (age of 5.0 Myr, PKP) with
such a low metallicity, Knot 3 is certainly one of the most attractive
targets for the future Extremely Large Telescopes (ELTs) to study individual
massive stars with
metallicities typical of very young galaxies in the early Universe.

\begin{figure}
  \centering
 \includegraphics[angle=-90,width=7.5cm, clip=]{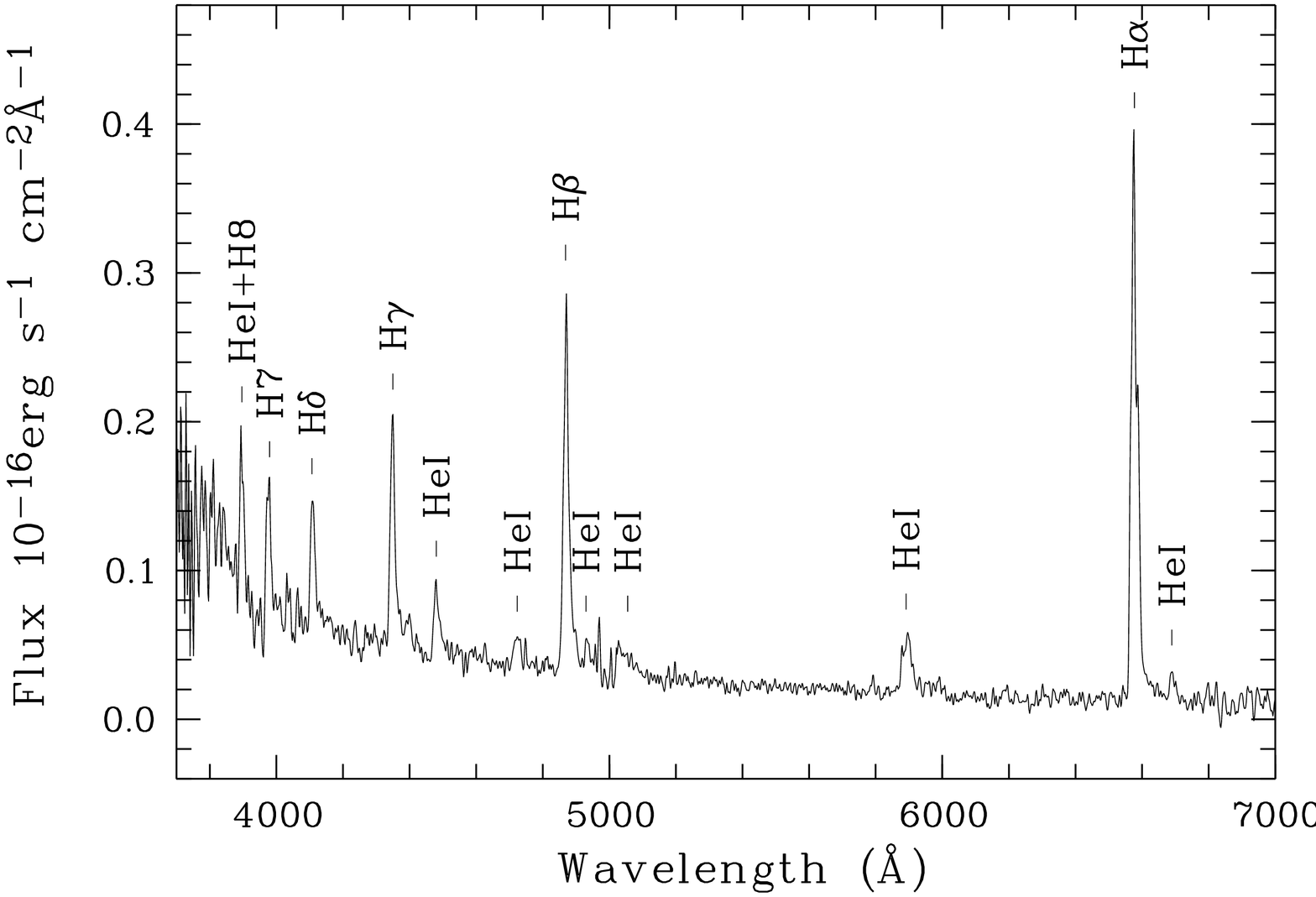}
 \includegraphics[angle=-90,width=7.5cm, clip=]{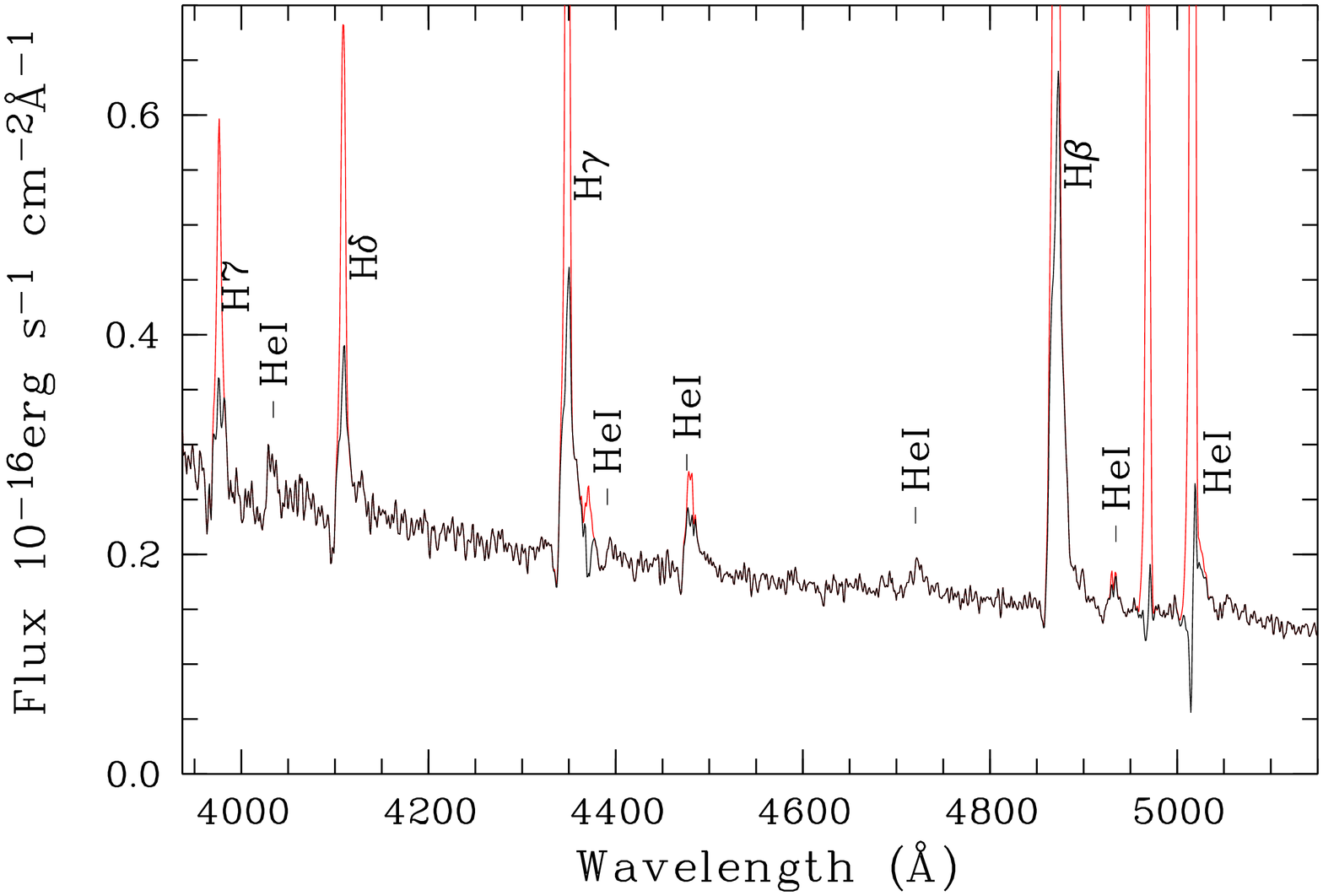}
  \caption{{\bf Top panel:} The difference spectrum of Knot~3 with
   resolution of $\sim$12~\AA, obtained by subtraction of the two spectra
   taken 3 years apart.
%   The old one was normalized by a factor of
%   0.85 to get the complete subtraction of [\ion{O}{iii}]$\lambda$4949,5007 lines.
  The residual features represent the massive star wind emission lines, while
  the underlying continuum presumably indicates the spectral energy
  distribution of a variable hot star. The marked lines % (in both panels)
  are H8+\ion{He}{i} 3889, H$\epsilon$ 3970, \ion{He}{i} 4026,
  H$\delta$ 4102, H$\gamma$ 4340, \ion{He}{i} 4388,
  \ion{He}{i} 4471, \ion{He}{i} 4713, H$\beta$ 4861, \ion{He}{i} 4922,
  \ion{He}{i} 5016 (affected by subtraction of [\ion{O}{iii}]5007),
  \ion{He}{i} 5876, H$\alpha$,  \ion{He}{i} 6678.
 {\bf Bottom panel:}
 The spectrum of Knot~3 with resolution of $\sim$6~\AA.
 Shown here are both the original spectrum (red) and a difference
 spectrum (black), obtained by the subtraction of scaled lines in the
 spectrum of Knot~1, registered on the same slit. Continuum remains original.
 Broad wings of both Balmer and \ion{He}{i} lines are visible as well as
 an indication of P~Cygni profiles in Balmer lines.
}
	\label{fig:dif_cont}
 \end{figure}

Fortunately, by good luck, some tracers of the individual massive stars
in nearby metal-poor galaxies can be found occasionally in the spectra of
star-forming regions \citep{ITG07}. This allows to conduct
their studies with modern large telescopes. DDO~68 appears to be one of
such galaxies, but outstanding for its the record lowest metallicity.
In Fig.~\ref{fig:knot3}, we display the old and new spectra of Knot~3, with
resolution of FWHM$\sim$12~\AA, in which pronounced differences have occurred
during the past 3 yr.
To better describe the variable component in the spectrum of Knot~3,
including its continuum, we show its difference spectrum in the top panel
of Fig.~\ref{fig:dif_cont}. The normalization of the old spectrum (by a
factor of 0.85) was performed before the subtraction, in order to fully
remove the (assumed) HII region nebular lines
[\ion{O}{iii}]$\lambda$4959,5007.

In the bottom panel of Fig.~\ref{fig:dif_cont}, we show the spectrum of Knot~3
with resolution of FWHM$\sim$6~\AA, in which broad components of Balmer and
He~I lines are more evident. The subtraction of nebular lines (shown in red)
in this spectrum was performed by scaling the same lines in the spectrum of
Knot~1, exposed on the same long slit, just $\sim$15~arcsec a side.
We present the parameters of `narrow' and `broad' components of the
residual lines, shown in black, in Table \ref{t:lines_residual}, where the
names of the columns are self-explaining. To determine parameters of `narrow'
and `broad' components (index $n$ and $b$ for Balmer and \ion{He}{i} lines) we
performed the procedure of two-component Gaussian fitting in MIDAS.
To understand the relative velocities of the transient emission and that
of Knot~3 HII region, the heliocentric velocity of the latter, defined
on the spectrum with FWHM=6~\AA, is V(Knot~3)=564$\pm$27~\kms.

\begin{table}
\centering{
\caption{Parameters of residual lines in Knot 3}
\label{t:lines_residual}
\begin{tabular}{lcccc} \hline  \hline
\rule{0pt}{10pt}
$\lambda_{\rm obs}$ & $\lambda_{0}$ Ion & V$_{\rm hel}$ & Flux$^a$  & FWHM \\
%\rule{0pt}{10pt}
%-------&----------------------&-----------------   -&------------  &---------
% Lobs  &$\lambda_{0}$ Ion     &V$_{\rm hel}$$\pm$err&Flux   Ferr   & FWHM   Err \\
 (\AA)  & (\AA)                &  \MC{1}{c}{(\kms)}  &              &\MC{1}{c}{(\kms)}  \\ \hline
%-------&----------------------&---------------------& -----------  &---------------
%\rule{0pt}{10pt}
4109.70 & 4101.74 H$\delta_n$  &   582$\pm$14        & 1.15$\pm$0.12&   711$\pm$34    \\
4126.82 & 4101.74 H$\delta_b$  &  1833$\pm$64        & 0.26$\pm$0.12&   728$\pm$159   \\
4349.27 & 4340.47 H$\gamma_n$  &   608$\pm$14        & 2.58$\pm$0.19&   803$\pm$39    \\
4361.94 & 4340.47 H$\gamma_b$  &  1483$\pm$386       & 0.84$\pm$0.53&  3182$\pm$663   \\
4479.51 & 4471.47 \ion{He}{i}$_n$      &   539$\pm$39        & 0.47$\pm$0.16&   732$\pm$114   \\
4492.80 & 4471.47 \ion{He}{i}$_b$      &  1430$\pm$345       & 0.41$\pm$0.23&  1440$\pm$558   \\
4723.16 & 4713.14 \ion{He}{i}$_n$      &   638$\pm$50        & 0.35$\pm$0.16&  1012$\pm$153   \\
4871.65 & 4861.33 H$\beta_n$   &   636$\pm$14        & 5.46$\pm$0.45&   805$\pm$42    \\
4887.94 & 4861.33 H$\beta_b$   &  1641$\pm$608       & 0.91$\pm$0.38&  1616$\pm$1007  \\
4932.97 & 4921.93 \ion{He}{i}$_n$      &   673$\pm$22        & 0.21$\pm$0.10&   605$\pm$52    \\
5023.99 & 5015.66 \ion{He}{i}$_n$      &   498$\pm$66        & 0.65$\pm$0.16&   777$\pm$145    \\
5054.28 & 5015.66 \ion{He}{i}$_b$      &  2308$\pm$410       & 0.29$\pm$0.29&  1517$\pm$1044  \\
\hline  \hline
\MC{5}{l}{$^a$ in units of 10$^{-16}$ ergs\ s$^{-1}$cm$^{-2}$.}\\
\end{tabular}
 }
\end{table}

\section[]{DISCUSSION}
\label{sec:dis}

The natural interpretation of this transient emission-line component is
the appearance of a luminous variable star.
Some of the observed and derived parameters of this unique variable star
and the related star-forming region are as follows. The integrated $V$-band
magnitude of Knot~3, as measured on the respective image in PKP,
is $\sim$21.2.
The latter is consistent with $V$ calculated through convolving the 2008
January spectrum of this region with the $V$-filter bandpass, namely
$V$=20.94.
%%%%  magnitudes with the transform formula from [13].
% Knot 5 BTA photom. V=19.40, SDSS V(g,r)=19.38
% Knot 3 SDSS V(g,r)=21.24
% Need to get V-mag of Knot 3 via convolution.
% Convolution of Knot 3 latest spectrum gives V=20.94
% V-mag of transient via convolution with V-band appears as 22.95, only
% 2.0 mag fainter.
A similar convolution of the residual spectrum in the top panel of
Fig.~\ref{fig:dif_cont} allows us to estimate the $V$-magnitude of the
transient to be 22.95. Its continuum, clearly rising to the ultraviolet (UV),
indicates a hot star of OB spectral type.
For the accepted distance module of DDO~68 $\mu$=30.0 (D=10 Mpc)
(see Pustilnik et al. 2008), the absolute magnitudes, M$_{\rm V}$, of Knot~3
and the variable star are --9.1 and --7.1, respectively.

For the estimated Knot~3 age of $\sim$5~Myr (PKP), one can calculate the
mass of its young stellar cluster, using, e.g., models for evolving stellar
populations of PEGASE2 package (Fioc \& Rocca-Volmerange 1999). Accounting
for $\sim$20 per cent light
contribution of nebular emission and a small correction for the Galaxy
extinction, with the accepted stellar metallicity of z=0.0004 and the standard
Salpeter Initial Mass Function with M$_{\rm low}$ and M$_{\rm up}$ of 0.1
and 120 M$_{\odot}$, we obtain M$_{\rm cluster}$ =
8.5$\times$10$^3$~M$_{\odot}$.

To further address the type of a massive star related to the observed
transient, we briefly describe its emission-line properties. In both spectra
of the variable massive star in Fig.~\ref{fig:dif_cont} only the Hydrogen
Balmer series and \ion{He}{i} emission lines are detected. This is in
agreement with its extremely low metallicity of Z$_{\odot}$/36.
The observed line profiles,
with broad asymmetric wings and an indication of P~Cygni profile
% (seen best in the higher resolution spectra in the bottom panel of
% Fig.~\ref{fig:dif_cont}),
resemble those of B-supergiant LBVs. These represent
the short transition phase in the
post-main-sequence evolution of massive stars with M$\lesssim$50 M$_{\odot}$.
The latter are exemplified by spectra of stars He3-519, AG~Car, $\eta$~Car
and others in Kudritzki (1998) and Walborn \& Fitzpatrick (2000). Their
`narrow' components are close in radial velocity
to that of the related HII region. However, their widths (FWHM) of
$\sim$700~\kms\ are clearly larger than the instrumental ones. The nature of
both `narrow' and broad components (FWHM$\sim$1000-2000~\kms) is not clear.
For the accepted DDO~68 distance of 10~Mpc, the H$\beta$ line fluxes
correspond to luminosities of $\sim$4.6 and $\sim$0.8 10$^{36}$ erg~s$^{-1}$
in `narrow' and broad components, respectively.
Both electron scattering in dense envelopes \citep{Bernat78}
and stellar winds could contribute to their appearance.
The asymmetry of the broad components can be partly due to absorption of the
approaching region of the (non-spherical) wind. This has important
implications, since in very low-metallicity stars the line-driven wind can
be inefficient, and continuum-driven eruptions might be more significant
\citep{SO2006}.
The latter, in turn, will have important implications for evolution of
Population III stars and the early metal enrichment of the Universe.

The hunting for the most metal-poor local galaxies as useful
testbeds for models of young galaxies was rather successful during the last
years, with the lowest HII-region oxygen abundance found so far in SBS
0335--052~W on the level of 12+$\log$(O/H)=7.12$\pm$0.03 \citep{ITG05}.
The star-forming region Knot~3 of a much closer galaxy DDO~68 appears
to have very similar metallicity of 12+$\log$(O/H)=7.10$\pm$0.06.
DDO~68 displays also other unusual properties, such as HI and optical
morphology and kinematics of a likely recent merger and a rather young
($\lesssim$1~Gyr) the oldest detectable stellar population (PKP),
\citep{Ekta08,DDO68_sdss}.

Due to its relative proximity and very low metallicity, this galaxy and its
young SF regions are among the best targets for detailed studies of
star-formation,
its feedback, and massive star properties typical of
high-redshift young objects. Monitoring of its several young
star-forming regions, with ages of 4 to 7 Myr, to discover variable massive
stars similar to the one reported here, will be very important. The
subsequent spectral studies of such objects, could lead to interesting
findings in extremely metal-poor environments and the properties of massive
metal-poor stars, actively studied in recent years
\citep[e.g.,][]{Meynet07,Yoon08}.
In particular, imaging with the HST of Knot~3 would help to localise this
luminous star ($V$$\sim$23), and provide its spectral energy distribution
in broad range from UV to NIR. This information will be helpful for
spectral study of this star with the next generation space telescope JWST.
Even its medium resolution spectra (R$\sim$5000) would provide the important
constraints on rotation of such stars, while NIR spectra might shed light on
the properties of related wind/shells.
%All these data would be a breakthrough,
%since no direct observations exists on massive stars at that low Z.

Massive low-metallicity stars are the main suppliers of energy, momentum
and metals during the early evolution of galaxies in the young Universe
\citep[e.g.,][]{Barkana01}. Their properties and details of
their evolution are important for estimating the effect of
SF feedback on galaxy evolution and their effect on
re-ionization of the Universe.
Thus, the discovery and the study of such metal-poor massive stars
with the next generation of ELTs is directly relevant to insights in the
early Universe evolution.

%Besides, the LBV phase, probably, can be in some objects the nearest
%in time to SN explosions (ref.??)  Therefore, one can catch the chance to
%witness SN explosion of very low-Z massive star, which in turn can be
%related to cosmological gamma-ray bursts (GRB).

%\section{Summary}
%\label{sec:summ}

Summarising the results and discussion above we draw the following
conclusions:

\begin{enumerate}
\item
The spectrum of HII-region Knot~3 in the nearby XMD galaxy DDO~68 changed
substantially during the period between 2005 January and 2008 January,
displaying enhanced intensities and broad components of Balmer and
\ion{He}{i} lines,
as well as an additional continuum component typical of a hot star.
\item
The luminosity, blue continuum and emission-line parameters of this variable
emission resemble those expected for a LBV with
metallicity of Z=Z\sunn/36 (as measured on nebular emission in Knot~3). This
suggests
the discovery of the first individual massive star with the record low
metallicity
and implies good prospects for study such stars with the future ELTs.
\end{enumerate}

\section*{Acknowledgements}
%{\bf ACKNOWLEDGMENTS} \\

SAP and ALT acknowledge the support of this work through the RFBR grant
No. 06-02-16617. The authors are grateful to A.~Moiseev for useful
information on Knot~3 related data, and to S.~Fabrika, O.~Sholukhova and
E.~Chentsov for valuable consultations on LBV stars. We thank Y.~Izotov
for useful discussion and providing with new information prior publication.
% and D.~Buckley for careful reading of the first version of paper.
Authors thank Y.~Balega for granting the BTA Director's Discretionary Time.
for additional spectroscopy of the DDO~68
transient.

%===========================================================================

\bsp

\label{lastpage}

\end{document}